\documentclass[11pt,a4paper]{article}
 \usepackage{amssymb,amsmath,amsfonts,amssymb}
 \usepackage{graphics,graphicx,color}
\newcommand{\s}[1]{{\textsf{\textbf{#1}}}}

\newcommand\rvec{{\bf r}}

\newcommand\xhat{{\bf \hat x}}
\newcommand\zhat{{\bf \hat z}}

\newcommand\evec{{\bf e}}

\newcommand\dvec{{\bf d}}

\newcommand\eps{{\epsilon}}

\newcommand{\Rr}{{\mathbb R}}

\title{\huge\s{Twisted rods, helices and buckling solutions in three dimensions}}
 \date{\today}
\author{{\Large Apala Majumdar and Alex Raisch}
\\  {\it Department of Mathematical Sciences, University of Bath}\\
            {E-mail:\ {\tt
majumdar@maths.ox.ac.uk}}\\
            { \it Corresponding author: A. M}. }

 \newtheorem{proposition}{Proposition}

\begin{document}

\maketitle

\section{Introduction}
\label{sec:intro}

The study of slender elastic structures is an archetypical problem in continuum 
mechanics, dynamical  systems and bifurcation theory, with a rich history dating
back to Euler's seminal work in the 18th century. These filamentary elastic 
structures have widespread applications in engineering and biology, examples 
of which include cables, textile industry, DNA experiments, collagen modelling 
etc~\cite{Leeetal2013, GoNiTa2001}. One is typically interested in the 
equilibrium configurations of these rod-like structures, their stability and
dynamic evolution and all three questions have been extensively addressed in
the literature, see for example~\cite{Maddocks1984, CaMa1984, Hoffman2004, NeuHen2004}
and more recently \cite{ReiPet2011, Manning2009, Manning2013}. However, it is 
generally recognized that there are still several open non-trivial questions related
to  three-dimensional analysis of rod equilibria, inclusion of topological and 
positional constraints and different kinds of boundary conditions.

In this paper, we study three different problems in the analysis of rod
equilibria ordered in terms of increasing complexity. All three problems focus 
on naturally straight, inextensible, unshearable rods with kinetic symmetry, 
subject to terminal loads and controlled end-rotation. The first problem is 
centered around the stability of the trivial solution or the unbuckled 
solution in three dimensions (3D), subject to a terminal load and controlled 
end-rotation with three different types of boundary conditions. This can be 
regarded as a generalization of the recent two-dimensional analysis of three
classical elastic strut problems in \cite{ReiPet2011}. We work with the 
Euler angle formulation for the rod geometry and work away from polar 
singularities; this excludes rods with self-intersection or self-contact 
but still accounts  for a large class of physically relevant configurations 
in an analytically tractable way \cite{Maddocks1984, MaPrGo2012, MaGo2013}.
We study the stability of the trivial solution for (i) purely Dirichlet 
conditions for the Euler angles, (ii) mixed Dirichlet-Neumann conditions for 
the Euler angles and (iii) purely Neumann conditions for Euler angles. We study
stability largely in terms of the positivity of the second variation of a 
rod energy within the Kirchhoff rod model, that is quadratic in the strain 
variables, as has been extensively used in the 
literature~\cite{Maddocks1984,CaMa1984,Manning2009, Manning2013}.  In the 
Dirichlet case, we optimize the stability estimates in our previous work 
in~\cite{MaPrGo2012}, i.e., we compute the critical force $F=F_c$ such that 
the trivial solution is locally stable, in the nonlinear sense, for all 
forces $F < F_c$ and unstable for $F > F_c$. We analytically compute 
eigenfunctions and the corresponding eigenvalues for the second variation 
operator, followed by bifurcation diagrams for the Euler angles in three 
dimensions, as a function of the applied load. The second variation analysis
only gives insight into local stability i.e. stability with respect to small
perturbations. In Proposition~\ref{prop:1}, we obtain global stability 
results for the trivial solution, in the presence of Dirichlet boundary 
conditions, valid for large perturbations within an explicitly defined class. 

Neumann boundary conditions pose new challenges for traditional methods 
in stability analysis \cite{Manning2009}. We analyze the stability of the 
trivial solution in 3D, subject to a terminal load and imposed twist, for 
Dirichlet-Neumann and Neumann-Neumann conditions. We do not appeal to 
conjugate-point type methods but directly work with the second variation 
of the rod-energy, without introducing boundary terms. The second variation 
is then simply a measure of the energy difference between the perturbed 
state and the trivial solution, for small perturbations, and we deduce 
conditions for the positivity of the energy difference leading to local 
stability. In particular, we obtain explicit stability estimates in terms 
of the material constants, applied force, twist and bypass the problems 
encountered in  conjugate-point type methods for Neumann boundary 
conditions~\cite{Manning2009}.

The second problem concerns the stability of helical rod equilibria in 3D,
subject to a terminal load in the vertical direction. We explicitly 
construct a family of helical equilibria of the rod-energy, following 
the methods in~\cite{ChGoMaJo2006}, for given values of the applied 
force and material/elastic constants. We then study the corresponding 
second variation of the rod-energy and analytically demonstrate that 
such non-trivial equilibria are stable for a range of compressive and 
tensile forces. In addition to a static stability analysis, we numerically 
study the dynamic evolution of these helices using a simple gradient-flow 
model for the rod-energy, based on the principle that the system evolves 
along a path of decreasing energy. Although, it would be more realistic to 
solve Kirchhoff nonlinear dynamic equations \cite{GoNiTa2001,NiGo1999}, 
we believe that the gradient flow model is simpler and yet preserves 
the qualitative features of the dynamic evolution. The stable helices, 
of course, remain stable with time but the unstable helices demonstrate 
interesting unwinding patterns, as a function of the imposed boundary 
conditions for the Euler angles.

The third problem focuses on the static and dynamic equilibria of the
localized buckling solutions reported in~\cite{NiGo1999}. We numerically 
find that these 
solutions are unstable, by computing negative eigenvalues of the Hessian 
operator associated with the quadratic rod-energy. We, further, adopt a 
gradient flow model for the dynamical evolution of these unstable equilibria 
and the evolution proceeds along a path of decreasing energy and reveals 
a variety of different spatio-temporal patterns, again strongly dependent
on the boundary conditions for the Euler angles. The numerical algorithm 
accounts for integral isoperimetric constraints on the evolution and can
be adapted to include a larger class of integral constraints. In all cases, 
we use a combination of variational techniques, analysis and numerical 
computations to study the static and dynamic stability of rod equilibria, 
ranging from the trivial solution to non-trivial helical solutions and 
finally buckled solutions. The analytical methods are transparent, simply 
rely on variational inequalities and are independent of any numerics, 
making them of independent interest for one-dimensional boundary-value 
problems. The numerics have  new features, see Section~\ref{sec:numerics}, 
are guided by the analysis and collectively, these methods offer new 
insight into the interplay of boundary conditions, terminal loads and 
stability in the analysis of rod equilibria.

The paper is organized as follows. In Section~\ref{sec:rod}, we review 
the Kirchhoff rod model; in Section~\ref{sec:straight}, we analyze three 
different boundary-value problems for the trivial unbuckled solution in 
3D. In Section~\ref{sec:helices}, we analytically and numerically study 
the static stability and dynamic evolution of prototypical helical 
equilibria in 3D and in Section~\ref{sec:goriely}, we focus on the 
localized buckling solutions reported in \cite{NiGo1999}.
In Section~\ref{sec:numerics}, we present a discretization of the 
gradient flow method and illustrate the incorporation of 
nonlinear integral constraints into the algorithm. In 
Section~\ref{sec:conclusions}, we outline our conclusions and 
future perspectives.

\section{The Kirchhoff Rod Model}
\label{sec:rod}

We work with a thin Kirchhoff rod whose geometry is fully
described by its centreline along with a frame that describes the
orientation of the material cross-section at each point of the
centreline~\cite{AnStKe1981, Antman2006, ChMa2004,GoNiTa2001,Maddocks1984}. 
We work in the thin filament approximation so that all physically 
relevant quantities are attached to the central axis and although this
approximation does not describe cross-sectional deformations, it
is suitable for long-scale geometrical and physical descriptions
of slender filamentary structures~\cite{ChGoMaJo2006,Maddocks1984}. 
The rod is inextensible, unshearable, uniform and isotropic by 
assumption~\cite{Maddocks1984, AnStKe1981}. Our methods can be 
generalized to extensible, anisotropic rods but we focus on simple
and generic cases to illustrate our methods \cite{MaPrGo2012}.

We denote the centreline by a space-curve, $\rvec(s)=\left( x(s),
y(s), z(s) \right):\Rr \to \Rr^3$, and the framing is described by
an orthonormal set of directors, $\left\{ \dvec_i(s) \right\}$,
$i=1,2,3$, where $s$ is the arc-length along the rod. In
particular, $\dvec_3$ is the tangent unit-vector to the rod axis
and inextensibility requires that \begin{equation} \label{eq:1}
\frac{ d \rvec}{ds} = \dvec_3,
\end{equation}
where $s \in \left[0, L \right]$ and $L$ is the fixed length of
the rod \cite{AnStKe1981, Antman2006,GoNiTa2001,ChGoMaJo2006,ChMa2004}. 
The orientation of the basis, $\left\{ \dvec_i(s) \right\}$, changes
smoothly relative to a fixed basis, $\left\{\evec_i \right\}$, 
and this change is described by
\begin{equation}
\label{eq:2} \frac{d \dvec_i}{ds} = \mathbf{\kappa} \times \dvec_i
\quad i=1,2,3
\end{equation}
where
$$ \mathbf{\kappa}  = \left( \kappa_1, \kappa_2, \kappa_3 \right)
$$
is the strain vector; $\kappa_1, \kappa_2$ contain information
about bending or curvature and $\kappa_3$ is the physical twist
\cite{MaPrGo2012, MaGo2013}.

We adopt the Euler angle formulation and use a set of Euler
angles, $\Theta(s) = \left( \theta(s), \phi(s), \psi(s) \right)$,
to describe the orientation of the director basis \cite{Maddocks1984,
MaPrGo2012}. The Euler angles are taken to be twice differentiable by
assumption, i.e., $\Theta \in C^2\left( [0, L]; \Rr^3 \right)$.
Further, we always take $0 < \theta < \pi$, i.e., we avoid the polar
singularities at $\theta=0$ and $\theta=\pi$ since a lot of our
mathematical machinery fails at the polar singularities
\cite{Maddocks1984}. This restriction necessarily excludes rods with
self-contact but still accounts for a large class of physically
relevant rod configurations, as we shall see in the subsequent
sections. The tangent vector, $\dvec_3$, is given by
\begin{eqnarray}
\label{eq:3b} \dvec_3 = \left( \sin\theta\cos\phi,
\sin\theta\sin\phi, \cos\theta \right) \end{eqnarray} and the rod
configuration can then be recovered from \eqref{eq:1}; there are
explicit expressions for $\dvec_1, \dvec_2$ but we do not need
them here \cite{Maddocks1984}. The strain components are given in
terms of the polar angles by
\begin{eqnarray} \label{eq:3} && \kappa_1 = -
\phi_s \sin\theta\cos\psi + \theta_s \sin\psi \nonumber\\
&& \kappa_2 = \phi_s \sin\theta \sin\psi + \theta_s \cos\psi
\nonumber \\ && \kappa_3 = \psi_s + \phi_s\cos\theta
\end{eqnarray}
where $\theta_s = \frac{d\theta}{ds}$ etc.

The isotropic rod, under consideration, obeys linear constitutive
stress-strain relations and has an isotropic, quadratic strain
energy given by
\begin{equation} \label{eq:4}
V[\theta,\phi,\psi]: = \int_{0}^{1} \frac{A}{2}\left( \theta_s^2 +
\phi_s^2 \sin^2\theta \right) + \frac{C}{2} \left(\psi_s +
\phi_s\cos\theta \right)^2 + \mathbf{F}L^2\cdot \dvec_3~ ds
\end{equation}
where the rod-length $L$ has been scaled away, $A, C >0$ are the
bending and twist elastic constants respectively with
$\frac{C}{A}\in \left[\frac{2}{3}, 1 \right]$ and $\mathbf{F}$
is an external terminal load \cite{NiGo1999}. We note that there 
are anisotropic rod energies with additional effects and constraints
in the literature but the simple energy in \eqref{eq:4} is regarded
as adequate for a large class of experiments in biology and
engineering \cite{Leeetal2013}. We are interested in modelling 
the rod equilibria, or equivalently the critical points of the
energy \eqref{eq:4} given by classical solutions of the associated
Euler-Lagrange equations:
\begin{eqnarray}
\label{eq:5} && A \theta_{ss} = A \phi_s^2\sin\theta\cos\theta - C
\phi_s\sin\theta(\psi_s + \phi_s\cos\theta) +
\frac{\partial}{\partial\theta} \left(\mathbf{F}L^2\cdot \dvec_3
\right) \nonumber \\ && \frac{d}{ds}\left[ A \phi_s \sin^2 \theta
+ C\cos\theta\left(\psi_s + \phi_s\cos\theta \right) \right] =
\frac{\partial}{\partial\phi} \left(\mathbf{F}L^2\cdot \dvec_3
\right) \nonumber
\\ && \psi_s + \phi_s \cos \theta = \Gamma
\end{eqnarray}
where $\Gamma$ is a constant for $0\leq s\leq 1$, depending on 
$(\mathbf{F},A,C,L)$. In what follows, we analyze the stability 
of prototypical rod equilibria under a variety of boundary 
conditions: Dirichlet conditions (clamped), Neumann conditions
(pinned), mixed conditions (Dirichlet for some Euler angles and 
Neumann for others), with and without isoperimetric constraints 
of the form
\begin{eqnarray}
\label{eq:6} \rvec_i (1) = \rvec_i (0)
\end{eqnarray}
for some $i=1,2,3$. If $\rvec(1)=\rvec(0)$ i.e. if
$\rvec_i(1)=\rvec_i(0)$ for all $i$, then we would have a closed
rod, which is outside the scope of this paper. The isoperimetric
constraints translate into integral constraints for the Euler
angles as shown below:
\begin{eqnarray}\label{eq:7}
&& x(1) = x(0) \Rightarrow \int_{0}^{1} \sin\theta\cos\phi~ds = 0
\nonumber \\ && y(1) = y(0) \Rightarrow
\int_{0}^{1}\sin\theta\sin\phi~ds = 0 \nonumber \\ && z(1) = z(0)
\Rightarrow \int_{0}^{1} \cos\theta~ds = 0.
\end{eqnarray}

\section{Boundary conditions and the stability of the ground
state} \label{sec:straight} 
\begin{figure}
\begin{center}
\includegraphics[scale = 0.2]{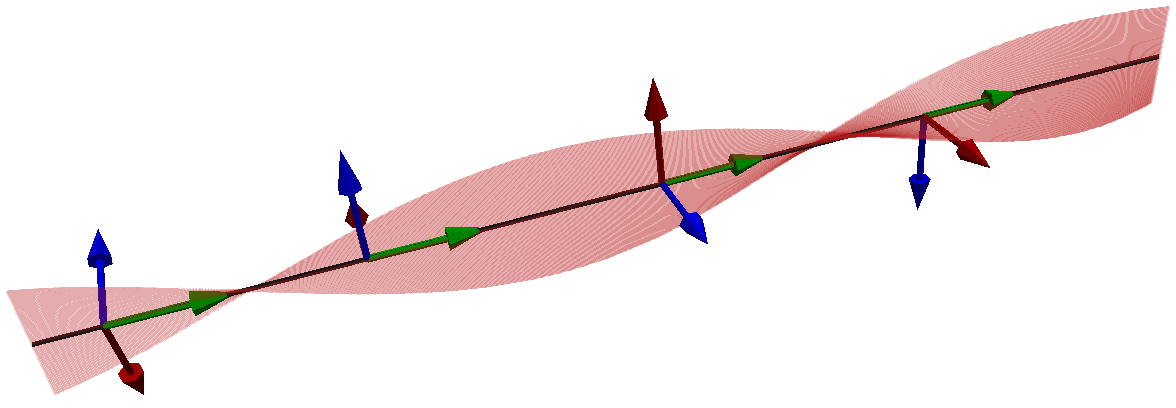}
\end{center}
\caption{Unbuckled twisted ground state with $L=1$ and $M=1$.
We plot the three directors $\mathbf{d}_1$ (red), 
$\mathbf{d}_2$ (blue) and the tangent $\mathbf{d}_3$ (green). 
Furthermore we emphasize the twist of the rod by a red 
ribbon which corresponds to $\mathbf{d}_1$.}
\label{P:unbuckled_ground}
\end{figure}

Our first example concerns a naturally
straight Kirchhoff rod, initially aligned along the $\xhat$-axis,
subject to controlled end-rotation and a terminal force,
$\mathbf{F}= F \xhat$, at the end $s=1$. This example builds on
our previous work in \cite{MaPrGo2012, MaGo2013} and our aim is 
to improve the previous results and carefully study the effect 
of boundary conditions on the stability of the unbuckled straight
state. A sample unbuckled ground state configuration can be seen
in Figure~\ref{P:unbuckled_ground}.

With $\mathbf{F}= F \xhat$, the rod-energy in \eqref{eq:4} becomes
\begin{eqnarray}\label{eq:8}
V[\theta,\phi,\psi]: = \int_{0}^{1} \frac{A}{2}\left( \theta_s^2 +
\phi_s^2 \sin^2\theta \right) + \frac{C}{2} \left(\psi_s +
\phi_s\cos\theta \right)^2 + FL^2 \sin\theta\cos\phi~ ds.
\end{eqnarray} In particular, $F>0$ corresponds to a compressive
force and $F<0$ corresponds to a tensile force.
It is trivial to note that the unbuckled state,
represented by the following triplet of Euler angles,
\begin{eqnarray}
\label{eq:9} \Theta_0(s) = \left( \frac{\pi}{2}, 0, 2\pi M s
\right)
\end{eqnarray}
is a solution of the Euler-Lagrange equations in \eqref{eq:5}. We
investigate the stability of $\Theta_0$ for three different
boundary-value problems \begin{eqnarray} \label{eq:10a}
\textrm{Dirichlet:} \Rightarrow \theta(0)=\theta(1)=\frac{\pi}{2}
\nonumber \\ \qquad \phi(0)=\phi(1) = 0 \nonumber \\ \qquad \qquad
\psi(0) = 0 \quad \psi(1) = 2\pi M,
\end{eqnarray}
\begin{eqnarray} \label{eq:10}
\textrm{Neumann:} \Rightarrow \theta_s(0)=\theta_s(1) = 0 \nonumber \\
\qquad \phi_s(0) = \phi_s(1) = 0 \nonumber \\  \qquad \qquad
\psi(0) = 0 \quad \psi(1) = 2\pi M,
\end{eqnarray} and
\begin{eqnarray} \label{eq:11}
\textrm{Mixed:} \Rightarrow \theta(0)=\theta(1) = \frac{\pi}{2} \nonumber \\
\qquad \phi_s(0) = \phi_s(1) = 0 \nonumber \\ \qquad \qquad
\psi(0) = 0 \quad \psi(1) = 2\pi M.
\end{eqnarray}

We first present some energy estimates that are useful for a
global and local stability analysis of $\Theta_0$. Our first
result concerns the energy difference between an arbitrary
configuration of Euler angles, $\Theta= (\theta,\phi, \psi)$, and
the unbuckled state, $\Theta_0 = \left(\frac{\pi}{2}, 0, 2\pi M s
\right)$. We can write $\Theta$ as
\begin{eqnarray}
\label{eq:12} &&\Theta(s) = \left( \theta(s) ,\phi(s),
\psi(s)\right) \nonumber \\ && \theta(s) = \frac{\pi}{2} +
\alpha(s) \quad 0\leq s\leq 1 \nonumber \\ && \phi(s) = \beta(s)
\quad 0\leq s \leq 1 \nonumber \\ && \psi(s) = 2\pi M s +
\gamma(s) \quad \gamma(0)=\gamma(1)=0,
\end{eqnarray}
since $\psi$ is subject to Dirichlet conditions in all three
boundary-value problems, \eqref{eq:10a}-\eqref{eq:11}. Further,
$-\frac{\pi}{2} < \alpha < \frac{\pi}{2}$ since $\theta$ does not
encounter the polar singularities by assumption. The functions,
$\alpha$ and $\beta$, measure the deviation of $(\theta,\phi)$
from the unbuckled state and are not subject to any end-point
constraints, since the choice of end-point constraints will depend
on the choice of the boundary-value problem in
\eqref{eq:10a}-\eqref{eq:11}.

From \eqref{eq:12}, it is straightforward to check that
\begin{equation}
\label{eq:13} \left( \psi_s + \phi_s \cos\theta \right)^2 = 4\pi^2
M^2 + \left(\gamma_s - \beta_s\sin\alpha \right)^2 + 4\pi M
\left(\gamma_s - \beta_s\sin\alpha \right).
\end{equation}
Therefore, using \eqref{eq:12}, \eqref{eq:13} and
$\gamma(0)=\gamma(1)=0$, we find that \begin{eqnarray}
\label{eq:14} && 2\left( V[\theta,\phi,\psi] -
V\left[\frac{\pi}{2}, 0, 2\pi M s\right] \right)= \nonumber \\ &&
= \int_{0}^{1} A \left(\beta_s^2\cos^2\alpha  + \alpha_s^2 \right)
+ C \left( \gamma_s - \beta_s \sin\alpha \right)^2 - 4 \pi M C
\beta_s\sin\alpha  + 2 F L^2\left(\cos\alpha \cos\beta - 1
\right)~ ds.
\end{eqnarray} Equation \eqref{eq:14} is valid for all triplets of
Euler angles provided they do not encounter the polar
singularities. Local stability analysis requires us to only focus
on small perturbations about $\Theta_0$. In this case, we consider
perturbations, $\Theta_\eps = \left(\theta_\eps, \phi_\eps,
\psi_\eps \right)$, where $0< \eps \ll 1$ is a small parameter and
\begin{eqnarray}
\label{eq:12b} &&\Theta_\eps(s) = \left( \theta_\eps(s)
,\phi_\eps(s), \psi_\eps(s)\right) \nonumber \\ && \theta_\eps(s)
= \frac{\pi}{2} + \eps \alpha(s) \quad 0\leq s\leq 1 \nonumber \\
&& \phi_\eps(s) = \eps \beta(s) \quad 0\leq s \leq 1 \nonumber \\
&& \psi_\eps(s) = 2\pi M s + \eps \gamma(s) \quad
\gamma(0)=\gamma(1)=0,
\end{eqnarray}
for all $0\leq s\leq 1$. Then, using Taylor expansions and
neglecting terms of order $\varepsilon^3$ and higher, 
\eqref{eq:14} simplifies to
\begin{eqnarray}
\label{eq:15b} && 2\left( V[\theta_\eps,\phi_\eps,\psi_\eps] 
- V\left[\frac{\pi}{2}, 0, 2\pi M s\right] \right) = \nonumber \\
&& = \eps^2 \int_{0}^{1}A\left(\alpha_s^2 + \beta_s^2 \right) 
+ C\gamma_s^2 - 4\pi M C \alpha \beta_s - FL^2\left(\alpha^2 
+ \beta^2 \right)~ds + O(\varepsilon^3).
\end{eqnarray} 
The second variation of the rod-energy about $\Theta_0$
is simply given by the right-hand side of \eqref{eq:15b} i.e.
\cite{MaPrGo2012,MaGo2013}
\begin{equation}
\label{eq:15}
\frac{d^2}{d\eps^2}V[\theta_\eps,\phi_\eps,\psi_\eps]|_{\eps=0}
=\int_{0}^{1}A\left(\alpha_s^2 + \beta_s^2 \right) + C\gamma_s^2 -
4\pi M C \alpha \beta_s - FL^2\left(\alpha^2 + \beta^2 \right)~ds.
\end{equation}

\subsection{Dirichlet problem}
\label{sec:dirichlet} We first consider the boundary-value problem
\eqref{eq:10a}. Then $\alpha$ and $\beta$ in \eqref{eq:12} must
vanish at the end-points. Whilst studying the static stability of
$\Theta_0$ under Dirichlet conditions, we frequently use 
Wirtinger's integral inequality cited below~\cite{Dacorogna2008,MaGo2013}.
\begin{proposition}
\label{prop:1} For every continuously differentiable function,
$u:[0,1] \to \Rr$ with $u(0)=u(1)=0$, we have
\begin{equation}
\label{eq:16} \int_{0}^{1}\left(\frac{d u}{ds} \right)^2~ds \geq
\pi^2 \int_{0}^{1} u^2(s)~ds.
\end{equation}
\end{proposition}


\begin{proposition}
\label{prop:3} The unbuckled state, $\Theta_0$, has lower energy
than all triplets of Euler angles, $\Theta=\left(\theta,\phi,\psi
\right)$, subject to the boundary conditions in \eqref{eq:10a},
provided that
\begin{eqnarray}
\label{eq:18}
&& \min_{s\in \left[0, 1 \right]} \cos^2 \alpha > \frac{4 M C}{A}
\nonumber \\ && {\rm max}\,\{FL^2,0\} 
< A\pi^2\min_{s\in \left[0, 1 \right]} \cos^2 \alpha -4\pi^2MC.
\end{eqnarray}
\end{proposition}

\textit{Proof:} We analyze the energy difference expression in
\eqref{eq:14}. Firstly, we note that
$$ \cos \alpha \geq 1 - \frac{\alpha^2}{2} $$
and hence
\begin{eqnarray}
\label{eq:19}  && F < 0 \Rightarrow 2 FL^2\left( \cos\alpha
\cos\beta - 1 \right) > 0 \nonumber \\
&& F > 0 \Rightarrow 2 FL^2\left( \cos\alpha \cos\beta - 1 \right)
\geq - FL^2 \left(\alpha^2 + \beta^2 \right) \qquad 0\leq s\leq 1.
\end{eqnarray} 
Secondly, we use Young's inequality and ${\rm sin}^2\alpha\leq\alpha^2$
to deduce that
\begin{equation}
\label{eq:20} \int_{0}^{1}\beta_s \sin\alpha~ ds \leq
\frac{1}{\delta}\int_{0}^{1}\alpha^2~ ds +
\delta\int_{0}^{1}\beta_s^2~ds
\end{equation}
for any positive real number $\delta$. We choose
$\delta=\frac{1}{\pi}$, substitute \eqref{eq:19} and \eqref{eq:20}
into \eqref{eq:14} to obtain
\begin{eqnarray}
\label{eq:21} && 2\left( V[\theta,\phi,\psi] -
V\left[\frac{\pi}{2}, 0, 2\pi M s\right] \right)\geq \nonumber
\\ && \geq \int_{0}^{1} A \left(\beta_s^2 \cos^2\alpha  + \alpha_s^2 \right) -
4\pi^2 M C \alpha^2 - 4 M C \beta_s^2 - \max\left\{F L^2,
0\right\}\left(\alpha^2 + \beta^2 \right)~ ds.
\end{eqnarray} Using Wirtinger's inequality \eqref{eq:16}, we
easily obtain
\begin{eqnarray}
\label{eq:22} && 2\left( V[\theta,\phi,\psi] -
V\left[\frac{\pi}{2}, 0, 2\pi M s\right] \right)\geq \nonumber\\
&& \left(  A \pi^2 \min_{s\in[0,1]}\cos^2\alpha - 4 \pi^2 M C -
\max\left\{F L^2,
0\right\} \right) \int_{0}^{1}\beta^2(s)~ds + \nonumber \\
&& + \left( A \pi^2 - 4\pi^2 M C - \max\left\{F L^2, 0\right\}
\right) \int_{0}^{1}\alpha^2~ ds.
\end{eqnarray} It is clear that
$$ V[\theta,\phi,\psi] -
V\left[\frac{\pi}{2}, 0, 2\pi M s\right] > 0 $$ if
\begin{eqnarray}
\label{eq:23} \max\left\{F L^2, 0\right\} < A \pi^2
\min_{s\in[0,1]}\cos^2\alpha - 4 \pi^2 M C.
\end{eqnarray} This completes the proof of
Proposition~\ref{prop:3}. $\Box$

Proposition~\ref{prop:3} effectively tells us that $\Theta_0$ is
the global energy minimizer, provided the twist ``M'' is
sufficiently small compared to the material constant,
$\frac{C}{A}$, and the Euler angles are bounded away from the
polar singularities. Proposition~\ref{prop:3} is a global result
whereas Proposition~\ref{prop:4} is a local stability result that
is an improvement over our previous results in \cite{MaPrGo2012}. We
recall that a rod equilibrium is stable in the static sense i.e.
is a local energy minimizer, if the second variation of the rod
energy is positive at the equilibrium \cite{Maddocks1984,Hestenes1966}. In
\cite{MaPrGo2012}, we show that $\Theta_0$ is stable in the static sense
for forces, $F < F_1$, where $F_1$ is an explicit expression in
terms of $M, C, A, L$. Correspondingly, we show that $\Theta_0$ is
unstable for forces, $F > F_2$, and $F_1 \neq F_2$. In
Proposition~\ref{prop:4}, we close the gap between the stability
and instability regimes.

\begin{proposition}\label{prop:4}
The unbuckled state, $\Theta_0$, is a local energy minimizer for
terminal forces
\begin{equation}
\label{eq:24} FL^2 < A\pi^2 \left( 1 - \left(\frac{M
C}{A}\right)^2 \right)
\end{equation} in \eqref{eq:8}. Correspondingly, $\Theta_0$ is unstable for
forces
\begin{equation}
\label{eq:25} FL^2 > A \pi^2 \left( 1 - \left(\frac{MC}{A}
\right)^2 \right).
\end{equation}
\end{proposition}

\textit{Proof:} An equilibrium, 
$\Theta^*=\left(\theta^*,\phi^*,\psi^*\right)$, 
is stable in the static sense if there exists a small 
neighbourhood~\cite{Maddocks1984,CaMa1984},
$$ \Delta(\Theta)= \left\{ \Theta=\left(\theta, \phi, \psi
\right): |\theta - \theta^*|^2 + |\phi^* - \phi|^2 + |\psi^* -
\psi|^2 \leq \eps \right\}, $$ such that
$$ V[\Theta] \geq V[\Theta^*] \qquad \forall \Theta \in \Delta.$$

We start by looking at the second variation of the rod-energy
evaluated at the unbuckled state, $\Theta_0$ in \eqref{eq:15} and
note that $\alpha$ and $\beta$ vanish at the end-points for the
Dirichlet boundary-value problem \eqref{eq:10a}. A simple
integration by parts shows that $\int_{0}^{1}\alpha \beta_s~ds = -
\int_{0}^{1} \beta \alpha_s~ds$ so that \eqref{eq:15} reduces to
\begin{equation}
\label{eq:26}
\frac{d^2}{d\eps^2}V[\theta_\eps,\phi_\eps,\psi_\eps]|_{\eps=0} =
\int_{0}^{1}A\left(\alpha_s^2 + \beta_s^2
\right) + C\gamma_s^2 - 2\pi M C \left(\alpha \beta_s - \beta
\alpha_s \right) - FL^2\left(\alpha^2 + \beta^2 \right)~ds.
\end{equation}
We write $\alpha$ and $\beta$ as
\begin{eqnarray}
\label{eq:27} \alpha = r \cos \sigma \nonumber \\
\beta = r \sin \sigma
\end{eqnarray}
with $r^2 = \alpha^2 + \beta^2$ and $r(0)=r(1)=0$. Straightforward
computations show that
\begin{eqnarray}
\label{eq:28}
\frac{d^2}{d\eps^2}V[\theta_\eps,\phi_\eps,\psi_\eps]|_{\eps=0}\geq
 \int_{0}^{1} A \left( r_s^2 + r^2 \sigma_s^2
\right) - 2\pi M C r^2 \sigma_s - FL^2 r^2~ds.
\end{eqnarray} It suffices to note that
$f(\sigma_s) = A \sigma_s^2 - 2 \pi M C \sigma_s \geq -
\left(\frac{\pi M C }{A}\right)^2 $ and the minimum is attained
for $\sigma_s = \frac{\pi M C}{A}$. Therefore,
\begin{equation}
\label{eq:29}
\frac{d^2}{d\eps^2}V[\theta_\eps,\phi_\eps,\psi_\eps]|_{\eps=0}\geq
 \int_{0}^{1} A r_s^2 - \left( \frac{\pi^2 M^2
C^2}{A} + FL^2 \right) r^2~ ds \geq
\int_{0}^{1} \left( A\pi^2 - \frac{\pi^2 M^2 C^2}{A} - FL^2
\right) r^2~ ds,
\end{equation}
wherein we have used Wirtinger's inequality,
$\int_{0}^{1}r_s^2 ds \geq \pi^2 \int_{0}^{1}r^2~ds. $ It is clear
that the second variation of the rod-energy in \eqref{eq:15} is
positive if
$$ FL^2 < A \pi^2 \left( 1 - \frac{M^2 C^2}{A^2}
 \right). $$

 Similarly, we can show that the second variation of the
 rod-energy in \eqref{eq:15}, about $\Theta_0$, is negative for
 $$ FL^2 > A \pi^2 \left( 1 - \frac{M^2 C^2}{A^2} \right),$$
by substituting
 \begin{eqnarray}
 \label{eq:30}
 && \alpha(s) = \sin \left(\pi s \right)\cos\left( \frac{\pi M C}{A}s \right)
 \nonumber \\ && \beta(s) = \sin \left(\pi s \right) \sin\left( \frac{\pi M
 C}{A} s \right)
 \end{eqnarray} in \eqref{eq:17}.
 The negativity of the second variation for a
 particular choice of $\left(\alpha, \beta \right)$ suffices to
 demonstrate the instability of $\Theta_0$ for forces $ FL^2 > A \pi^2 \left( 1 - \frac{M^2
 C^2}{A^2} \right)$ \cite{Maddocks1984,MaPrGo2012}. This completes the proof of
 Proposition~\ref{prop:4}. $\Box$

 \subsubsection{Bifurcations from $\Theta_0$}
 \label{sec:bifurcation}
 The local stability analysis in Proposition~\ref{prop:4} relies
 on the integral expression for the second variation of the
 rod-energy in \eqref{eq:15} and simple integral inequalities.
 Conjugate-point methods are an alternative and very successful
 approach to stability analysis; see \cite{Manning2013, Hoffman2004}.
 Here, we present a conjugate-point method type analysis for the
 unbuckled state, $\Theta_0$, in three dimensions and compute
 bifurcation diagrams for the Euler angles, $(\theta,\phi)$.

 We can use integration by parts to write the second variation in
 \eqref{eq:15} as
 \begin{equation}
 \label{eq:31}
 \frac{d^2}{d\eps^2}V[\theta_\eps,\phi_\eps,\psi_\eps]|_{\eps=0} 
 = \int_{0}^{1} \left(\alpha, \beta \right) \cdot 
 S\left(\alpha, \beta \right)~ds + \int_{0}^{1} C \gamma_s^2
\end{equation}
where $S(\alpha, \beta)$ is a coupled system of two linear
ordinary differential equations as shown below:
\begin{eqnarray}
\label{eq:32} && S(\alpha) = - A \alpha_{ss} - 2\pi M C \beta_s -
FL^2\alpha \nonumber \\ && S(\beta) = -A \beta_{ss} + 2\pi M C
\alpha_s - F L^2\beta.
\end{eqnarray}
From standard results in spectral theory \cite{Manning2013}, every
admissible $\left(\alpha, \beta \right)$ subject to $\alpha(0) =
\alpha(1)=0$ and $\beta(0)=\beta(1)=0$ can be written as a linear
combination of the eigenvectors of the second-order differential
operator, $S(\alpha, \beta)$, in \eqref{eq:31}. One can check that
there are two families of orthogonal eigenfunctions for $(\alpha,
\beta)$ given by
\begin{eqnarray}
\label{eq:33} && \left(\alpha, \beta \right)_m = \sin m \pi s
\left( \cos \frac{\pi M C}{A} s, \sin \frac{\pi M C}{A} s \right)
\quad m \in \mathbb{N} \nonumber
\\ && \left(\alpha, \beta \right)_m
= \sin m \pi s \left( -\sin \frac{\pi M C}{A}s, \cos 
\frac{\pi M C}{A} s \right) \quad m \in \mathbb{N},
\end{eqnarray} with corresponding eigenvalues
\begin{equation}
\label{eq:34} \lambda_m = A m^2 \pi^2 - \frac{\pi^2 M^2 C^2}{A} -
FL^2.
\end{equation} Equation~\eqref{eq:34} allows us to track the index
or equivalently the number of negative eigenvalues as a function
of the applied force and identify the critical forces
\begin{equation}
\label{eq:35} F_m = \frac{1}{L^2}\left( A m^2 \pi^2 - \frac{\pi^2
M^2 C^2}{A} \right).
\end{equation}

Hence, every $(\alpha, \beta)$ can be written as
\begin{eqnarray}
\label{eq:36} \left(\alpha, \beta \right) = \sum_{m=1}^{\infty}
a_m \sin m \pi s \left( \cos \frac{\pi M C}{A} s, \sin \frac{\pi M
C}{A} s \right) + \sum_{m=1}^{\infty} b_m \sin m \pi s \left(
-\sin \frac{\pi M C}{A} s, \cos \frac{\pi M C}{A} s \right)
\end{eqnarray} and substituting \eqref{eq:36} into \eqref{eq:31},
we obtain
\begin{eqnarray}
\label{eq:37} &&
\frac{d^2}{d\eps^2}V[\theta_\eps,\phi_\eps,\psi_\eps]|_{\eps=0} =
\int_{0}^{1} \left(\alpha, \beta \right) \cdot S\left(\alpha,
\beta \right)~ds + \int_{0}^{1} C \gamma_s^2 \geq \nonumber \\ &&
\geq \frac{1}{2}\sum_{m=1}^{\infty} \lambda_m \left(a_m^2 + b_m^2
\right).
\end{eqnarray}
From \eqref{eq:34}, it is clear that the smallest eigenvalue
satisfies $\lambda_1 > 0$ for forces 
$FL^2 < A \pi^2 - \frac{\pi^2 M^2 C^2}{A}$ and thus, the second 
variation of the rod-energy in \eqref{eq:15} is strictly positive 
for forces $FL^2 < A \pi^2 - \frac{\pi^2 M^2 C^2}{A}$.

\begin{figure}
\begin{center}
\includegraphics[scale = 0.205]{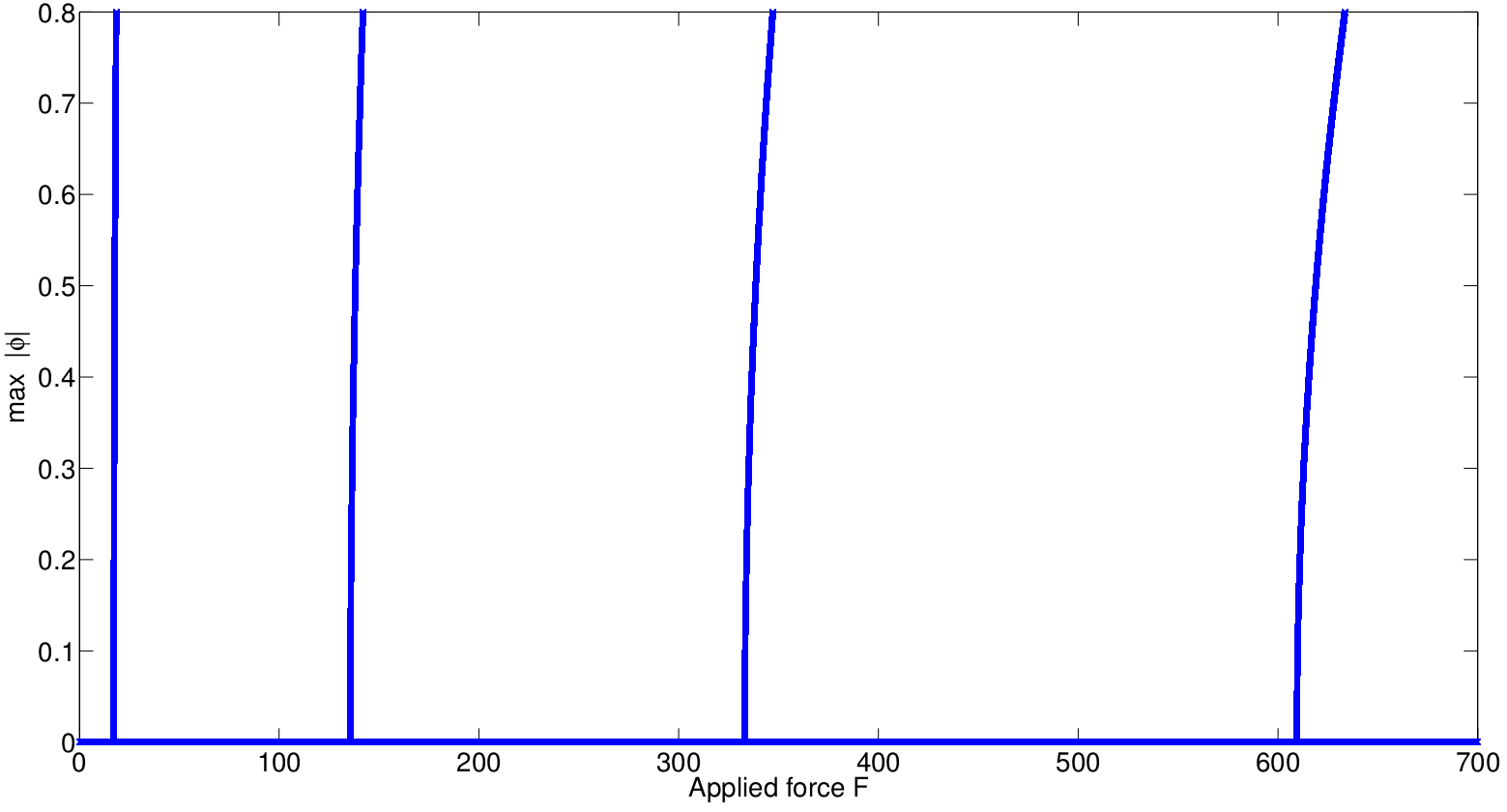}
\includegraphics[scale = 0.205]{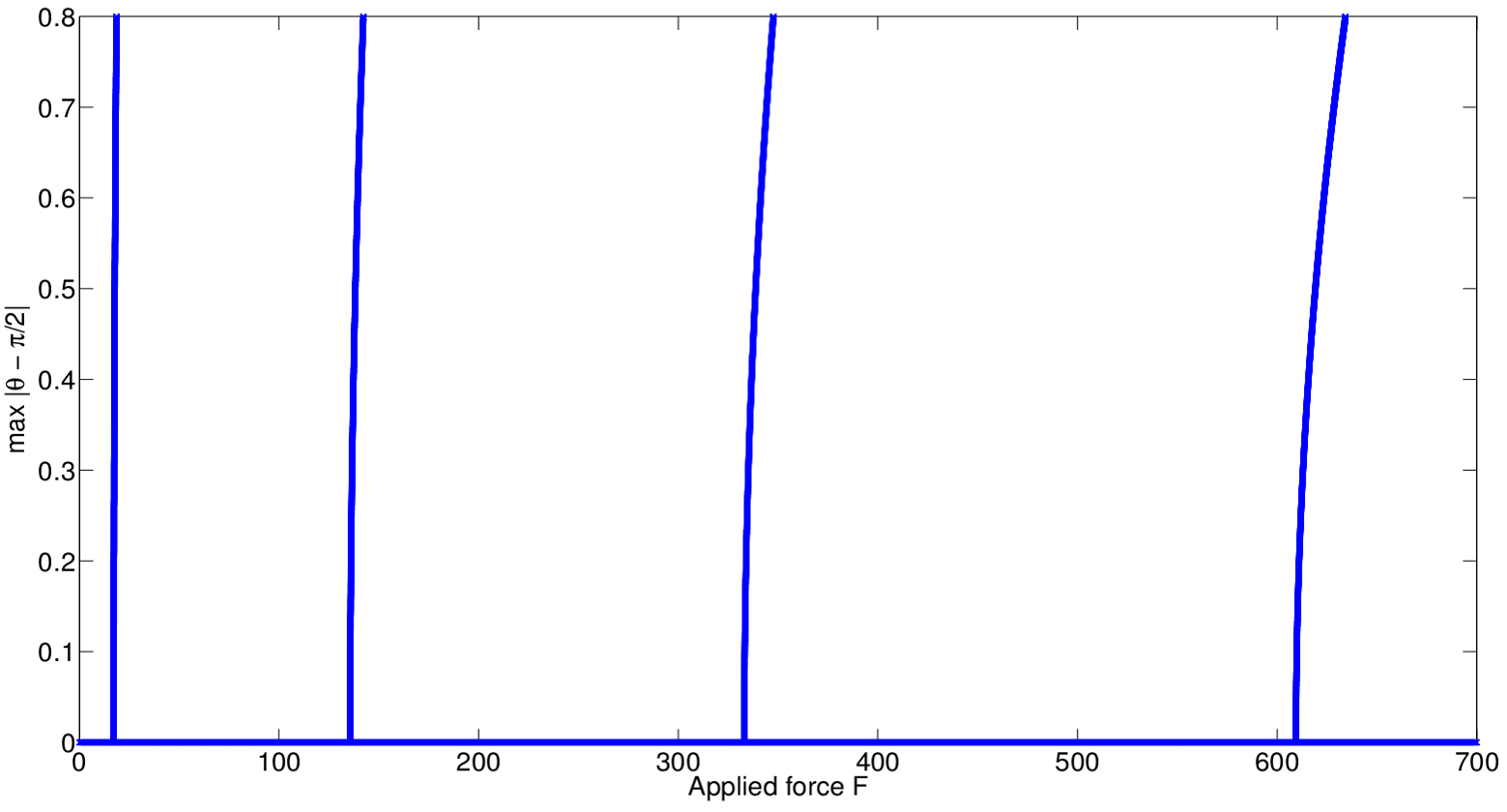}
\end{center}
\caption{Bifurcation plot for $\theta$ and $\phi$ with branches 
at $m=1,\dots,4$. The parameter setting is $C/A=3/4$, $M=1$ and $L=1$.}
\label{P:bif_plot}
\end{figure}

In Figure~\ref{P:bif_plot}, we plot bifurcation diagrams for the Euler
angles $(\theta,\phi)$ from the trivial solution, $\Theta_0$, as a
function of the applied load, $FL^2$. As can be seen from
Figure~\ref{P:bif_plot}, there is a bifurcating branch at every
critical force, $F_m$, with $m\geq 1$, and the bifurcation
diagrams are qualitatively similar to the well-known bifurcation
diagrams for the polar angle, $\theta$, in two dimensions
\cite{Manning2013, Manning2009}.

\begin{figure}
\begin{center}
\includegraphics[scale = 0.5]{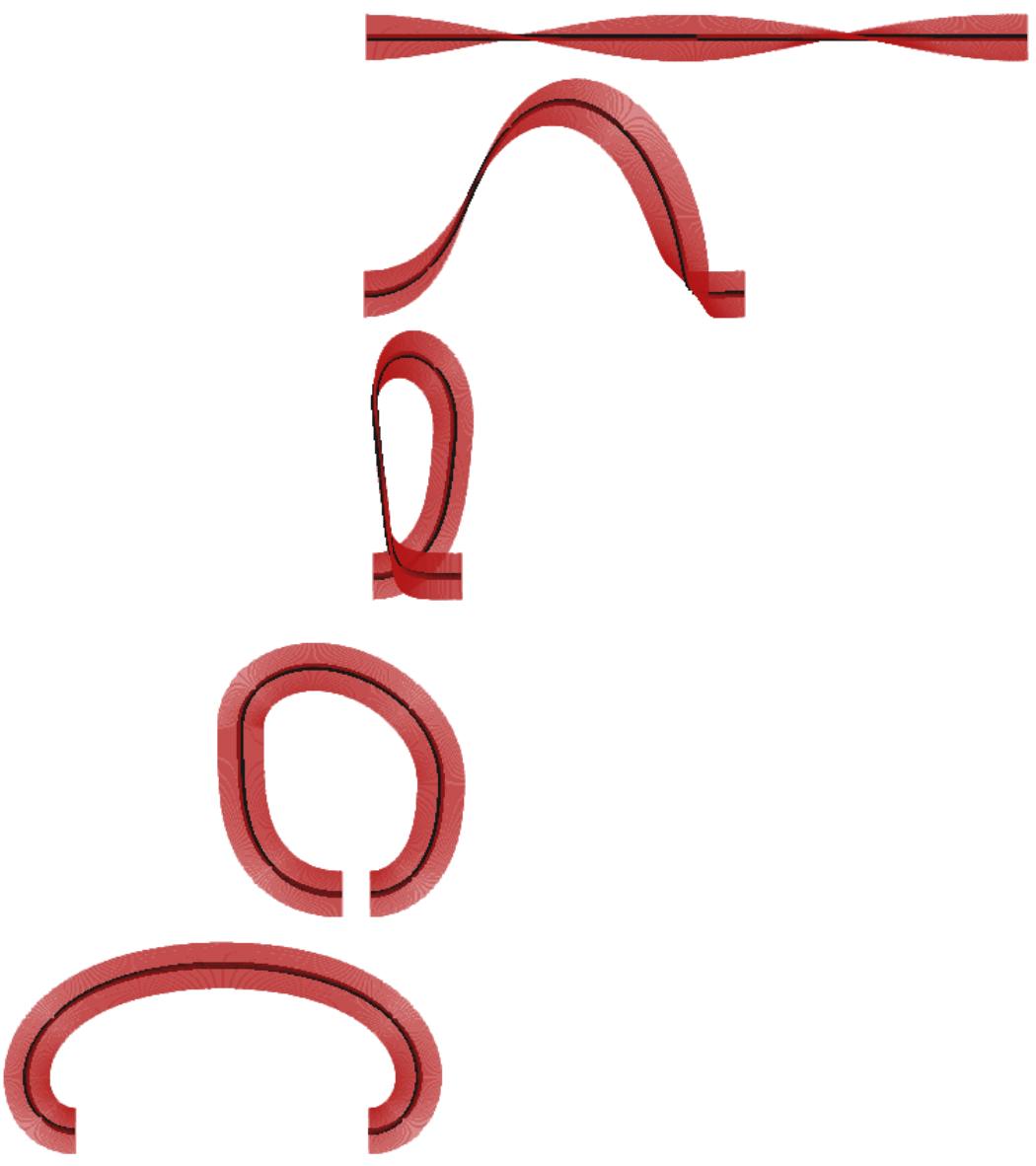}
\end{center}
\caption{Evolution of the unbuckled state under a $L^2$-gradient
flow. Parameters are $L=1$, $M=1$ and $C/A=1$. The applied 
force is $F=(50,0,0)$. Isoperimetric constraints ensure that
$y(1)=y(0)$ and $z(0)=z(1)$.}
\label{P:unbuckled_state}
\end{figure}

\subsubsection{Remarks on isoperimetric constraints}
\label{sec:iso} The local stability analysis in
Section~\ref{sec:dirichlet} can be generalized to the
boundary-value problem \eqref{eq:10a} augmented with the following
isoperimetric constraints:
\begin{eqnarray}
\label{eq:38} && y(0)=y(1) \Rightarrow \int_{0}^{1}\sin\theta \sin
\phi~ds = 0 \nonumber \\ && z(0)=z(1) \Rightarrow
\int_{0}^{1}\cos\theta~ds=0.
\end{eqnarray}
We consider ``small'' perturbations about $\Theta_0$ in
\eqref{eq:12}, as in \eqref{eq:12b}. The isoperimetric constraints
\eqref{eq:38} translate into the following integral constraints
for $\alpha, \beta$ for small perturbations,
\begin{eqnarray}
\label{eq:39} && \int_{0}^{1} \alpha (s) ds = 0 \nonumber \\
&& \int_{0}^{1}\beta(s) ds = 0.
\end{eqnarray}
The problem of local stability analysis of $\Theta_0$ subject to
\eqref{eq:38} reduces to a study of the second variation of the
rod-energy in \eqref{eq:15}, subject to the integral constraints
\eqref{eq:39}.
\begin{proposition}
\label{prop:5} The unbuckled state, $\Theta_0$, is stable in the
static sense for the boundary-value problem \eqref{eq:10a} and the
isoperimetric constraints \eqref{eq:38} for forces
\begin{equation}
\label{eq:40} FL^2 < A\pi^2 - \frac{\pi^2 M^2 C^2}{A}.
\end{equation}
Correspondingly, $\Theta_0$ is an unstable equilibrium of the
rod-energy, subject to the boundary conditions in \eqref{eq:10a}
and the constraints in \eqref{eq:38} for forces
\begin{equation}
\label{eq:41a} F L^2 > A\pi^2\left(1 - \left(\frac{MC}{A}\right)^2\right),
\end{equation}
if
\begin{equation}
\label{eq:41} \frac{MC}{A} = 2 p + 1 \quad p\in \mathbb{N}; p\geq
1.
\end{equation}
\end{proposition}

\textit{Proof:} It is trivial to check that $\Theta_0$ satisfies
the boundary conditions in \eqref{eq:10a} and the integral
constraints \eqref{eq:38}. The proof of local stability for forces
satisfying \eqref{eq:40}, is identical to the proof of
Proposition~\ref{prop:4}. In contrast to Proposition~\ref{prop:4},
we cannot prove instability of $\Theta_0$ in the complementary
regime with the constraints \eqref{eq:39}, for all values of
$\frac{MC}{A}$.

For $\frac{MC}{A}$ satisfying \eqref{eq:41}, one can check that
\begin{eqnarray}
\label{eq:42} && \alpha^*(s) = \sin \left(\pi s \right)\cos
\left(\frac{\pi M C}{A}s \right) \nonumber \\ 
&& \beta^*(s) = \sin \left(\pi s \right) \sin
\left( \frac{\pi M C}{A} s \right)
 \end{eqnarray} satisfy the integral constraints \eqref{eq:39}.
 Thus $(\alpha^*, \beta^*)$ qualify as an admissible perturbation
 that vanish at the end-points, $s=0$ and $s=1$, and satisfy the
 integral constraints \eqref{eq:39}. We substitute $(\alpha^*,
 \beta^*)$ into \eqref{eq:15} with $\gamma=0$ and find that
 \begin{equation}
 \label{eq:43}
 \frac{d^2}{d\eps^2}V[\theta_\eps,\phi_\eps,\psi_\eps]|_{\eps=0} <
 0
 \end{equation}
 for forces 
 \[
  F L^2 > A\pi^2\left(1 - \left(\frac{MC}{A}\right)^2\right),
 \]
 completing the proof of Proposition~\ref{prop:5}. $\Box$

 \subsection{Neumann boundary conditions}
 \label{sec:neumann}

 We analyze the local stability of $\Theta_0$ subject to the
 Neumann conditions \eqref{eq:10}. We consider small
 perturbations,
 as in \eqref{eq:12b}, and study the second variation of the
 rod-energy in \eqref{eq:15},
 \begin{equation}
\label{eq:15c}
\frac{d^2}{d\eps^2}V[\theta_\eps,\phi_\eps,\psi_\eps]|_{\eps=0}
=\int_{0}^{1}A\left(\alpha_s^2 + \beta_s^2 \right) + C\gamma_s^2 -
4\pi M C \alpha \beta_s - FL^2\left(\alpha^2 + \beta^2 \right)~ds,
\end{equation}
 subject to
 \begin{eqnarray}
 \label{eq:44}
 && \alpha_s(0) = \alpha_s(1) = 0 \nonumber \\
 && \beta_s(0) = \beta_s(1) = 0.
 \end{eqnarray}

 \begin{proposition}
 \label{prop:6}
 The unbuckled state, $\Theta_0$, is a locally stable equilibrium
 of the rod-energy in \eqref{eq:4}, subject to the boundary
 conditions in \eqref{eq:10}, for forces
 \begin{eqnarray}
 \label{eq:45}
 && FL^2 < -4 \frac{\pi^2 M^2 C^2}{A} \quad \textrm{for $A < 2 M
 C$} \nonumber \\
 && FL^2 < - 2\pi^2 M C \quad \textrm{for $A > 2 M C$.}
 \end{eqnarray}
 \end{proposition}

 \textit{Proof:} For $A < 2 M C$, we write the second variation in
 \eqref{eq:15} as
 \begin{equation}
 \label{eq:46}
 \frac{d^2}{d\eps^2}V[\theta_\eps,\phi_\eps,\psi_\eps]|_{\eps=0}
=\int_{0}^{1} A \left(\frac{2 \pi M C}{A} \alpha - \beta_s
\right)^2 + C \gamma_s^2 + A \alpha_s^2 - FL^2\beta^2 - \left(
FL^2 + \frac{4 \pi^2 M^2 C^2}{A} \right)\alpha^2~ds
\end{equation}
and it is clear that
$$
\frac{d^2}{d\eps^2}V[\theta_\eps,\phi_\eps,\psi_\eps]|_{\eps=0}>
0$$ for $$FL^2 + \frac{4 \pi^2 M^2 C^2}{A}  < 0 $$ as stated in
\eqref{eq:45}.

For $A> 2MC$, we write the second variation in
 \eqref{eq:15} as
\begin{equation}
 \label{eq:47}
 \frac{d^2}{d\eps^2}V[\theta_\eps,\phi_\eps,\psi_\eps]|_{\eps=0}
=\int_{0}^{1}  2MC\left(\pi \alpha - \beta_s \right)^2 + C
\gamma_s^2 + A\alpha_s^2 + \left(A - 2MC\right) \beta_s^2 -
FL^2\beta^2 -\left(FL^2 + 2\pi^2 M C \right)\alpha^2 ~ds
\end{equation} and it is clear that $\frac{d^2}{d\eps^2}V[\theta_\eps,\phi_\eps,\psi_\eps]|_{\eps=0}>
0$ for
$$ FL^2 + 2\pi^2 M C < 0 $$
as stated in \eqref{eq:45}. This completes the proof of
Proposition~\ref{prop:6}. $\Box$

\subsection{Mixed boundary conditions}
 \label{sec:mixed}

 We analyze the local stability of $\Theta_0$ subject to the
 mixed boundary conditions in \eqref{eq:11} i.e. Dirichlet for $\theta$ and Neumann for $\phi$.
 We consider small
 perturbations,
 as in \eqref{eq:12b}, and study the second variation of the
 rod-energy in \eqref{eq:15},
 \begin{equation}
\label{eq:15r}
\frac{d^2}{d\eps^2}V[\theta_\eps,\phi_\eps,\psi_\eps]|_{\eps=0}
=\int_{0}^{1}A\left(\alpha_s^2 + \beta_s^2 \right) + C\gamma_s^2 -
4\pi M C \alpha \beta_s - FL^2\left(\alpha^2 + \beta^2 \right)~ds,
\end{equation}
 subject to
 \begin{eqnarray}
 \label{eq:48}
 && \alpha(0) = \alpha(1) = 0 \nonumber \\
 && \beta_s(0) = \beta_s(1) = 0.
 \end{eqnarray}
In particular, we can use Wirtinger's inequality for $\alpha$ i.e.
$\int_{0}^{1}\alpha_s^2 ds \geq \pi^2 \int_{0}^{1}\alpha^2 ds$.
Note, that the same calculations can be done if the roles of 
$\alpha$ and $\beta$ in~\eqref{eq:48} are changed.

\begin{proposition}
 \label{prop:7}
 The unbuckled state, $\Theta_0$, is a locally stable equilibrium
 of the rod-energy in \eqref{eq:4}, subject to the boundary
 conditions in \eqref{eq:11}, for forces
 \begin{eqnarray}
 \label{eq:49}
 && FL^2 < \min\left\{ \pi^2\left(A - \frac{4 M^2 C^2}{A} \right), 0 \right\}
 \end{eqnarray}
 and unstable for forces
 \begin{eqnarray} \label{eq:50}
 && FL^2 > A\pi^2\frac{1-\Lambda^2}{1+\Lambda^2},
 \end{eqnarray}
where $\Lambda:=\frac{2MC}{A}$. Note, that for $A = 2MC$ we have a sharp
result, i.e., $\Theta_0$ is stable for $F<0$ and unstable for $F>0$.
 \end{proposition}

 \textit{Proof:} We write the second variation of the rod-energy
 about $\Theta_0$ as in \eqref{eq:46}.
 \begin{eqnarray}
 \label{eq:51} &&
 \frac{d^2}{d\eps^2}V[\theta_\eps,\phi_\eps,\psi_\eps]|_{\eps=0}
=\int_{0}^{1} A \left(\frac{2 \pi M C}{A} \alpha - \beta_s
\right)^2 + C \gamma_s^2 + A \alpha_s^2 - FL^2\beta^2 - \left(
FL^2 + \frac{4 \pi^2 M^2 C^2}{A} \right)\alpha^2~ds \geq \nonumber
\\ && \geq \int_{0}^{1} A \left(\frac{2 \pi M C}{A} \alpha - \beta_s
\right)^2 + C \gamma_s^2  - FL^2\beta^2 +\left( A \pi^2 -  \left(
FL^2 + \frac{4 \pi^2 M^2 C^2}{A} \right)\right)\alpha^2~ds,
\end{eqnarray} where we have used Wirtinger's inequality in the
second step. It follows immediately from \eqref{eq:51} that
$$\frac{d^2}{d\eps^2}V[\theta_\eps,\phi_\eps,\psi_\eps]|_{\eps=0}
> 0 $$
if $$FL^2 < \min\left\{ \pi^2\left(A - \frac{4 M^2 C^2}{A}
\right), 0 \right\}, $$ as stated in \eqref{eq:49}.

Similarly, we substitute
\begin{eqnarray}
\label{eq:52} 
&& \alpha^*(s) = \sin \pi s \nonumber \\
&& \beta^*(s) = -\frac{2MC}{A}\cos\pi s
\end{eqnarray}
into \eqref{eq:51} and find that
\[
 \frac{d^2}{d\eps^2}V[\theta_\eps,\phi_\eps,\psi_\eps]|_{\eps=0} = 
 -\frac{FL^2}{2}\left(1+\left(\frac{2MC}{A}\right)^2\right) + 
 \frac{A\pi^2}{2}\left(1-\left(\frac{2MC}{A}\right)^2\right)
\]

for this particular choice of $(\alpha^*, \beta^*)$. Therefore,
we conclude that the second variation is negative, if

\[
 FL^2 > A\pi^2\frac{1-\Lambda^2}{1+\Lambda^2}. 
\]
This completes the proof of Proposition~\ref{prop:7}. $\Box$

\section{Model helices}
\label{sec:helices} We construct prototype helical
equilibria for a naturally straight, inextensible, unshearable rod
that is subject to a terminal load, $\mathbf{F}=F \zhat$. This
choice of terminal load is motivated by the DNA manipulation
experiments reported in the literature \cite{FaRuOs1997}. 
The rod-energy is then given by
\begin{equation}
\label{eq:53} V[\theta,\phi,\psi]:=\int_{0}^{1}\frac{A}{2}
\left(\theta_s^2 + \phi_s^2 \sin^2\theta \right) + \frac{C}{2}
\left(\psi_s + \phi_s\cos\theta \right)^2 + FL^2 \cos \theta~ ds,
\end{equation} where $L$ is the fixed length of the rod. The 
corresponding Euler-Lagrange equations are
given by
\begin{eqnarray}
\label{eq:54}
&& A\theta_{ss} = A \phi_s^2 \sin\theta\cos\theta 
- C \phi_s \sin \theta (\psi_s + \phi_s\cos \theta) 
- FL^2 \sin \theta \nonumber \\
&& \frac{d}{ds}\left[ A \phi_s \sin^2\theta 
+ C \cos\theta\left(\psi_s + \phi_s\cos \theta \right) \right] 
= 0 \nonumber \\
&& \psi_s + \phi_s \cos \theta = \Gamma
\end{eqnarray}
where $\Gamma$ is a constant that depends on 
$\left(F, A, C\right)$..
It is straightforward to check that for given values of 
$\left(F,C,A, \lambda, \theta_0 \right)$, the following family of
 rod configurations,
 $\Theta_\lambda = \left(\theta_\lambda, \phi_\lambda, \psi_\lambda \right)$, 
 given by
\begin{eqnarray}
\label{eq:55}
&& \theta_\lambda(s) = \theta_0 \quad 0<\theta_0<\pi \nonumber \\
&& \phi_\lambda(s) = 2\pi \lambda s  \quad 0\leq s\leq 1 \nonumber \\
&& \psi_\lambda(s) = \left[ 2\pi\lambda\cos \theta_0\left( \frac{A}{C}-1 \right)
- \frac{FL^2}{2\pi \lambda C} \right] s + \xi
\end{eqnarray}
for any real number $\xi \in \Rr$, are exact solutions of the
Euler-Lagrange equations \eqref{eq:54}, subject to their own boundary
 conditions. We take $\theta_0 \in (0, \pi)$ so that we do not 
 encounter the polar singularities \cite{MaPrGo2012, Maddocks1984}.
 We note that the twist depends on the applied force i.e. for a 
 given set of parameters, $\left(F,C,A, \lambda, \theta_0 \right)$,
 the twist is given by
\begin{equation}
\label{eq:twist}
\psi_s + \phi_s \cos \theta = \frac{2 A}{C} \pi \lambda \cos \theta_0
- \frac{F L^2}{2 \pi \lambda C}.
\end{equation}

The solutions, $\Theta_\lambda$, are helices with constant curvature,
$\kappa$, and constant torsion, $\eta$, given by \cite{ChGoMaJo2006}
\begin{eqnarray}
\label{eq:56}
&& \kappa = 2 \pi \lambda \sin \theta_0 \nonumber\\
&& \eta = 2\pi \lambda \cos \theta_0.
\end{eqnarray}

The next step is to investigate the stability of the helical equilibria, 
$\Theta_\lambda$ in \eqref{eq:55}, subject to its own boundary conditions. 
For each $\lambda \in \Rr$ and $\theta_0 \in (0, \pi)$, we define the 
following Dirichlet problem for the Euler angles:
\begin{eqnarray}\label{eq:57}
&& \theta(0) = \theta(1) = \theta_0 \quad 0< \theta_0 < \pi \nonumber \\
&& \phi(0) = 0, \quad \phi(1)=2\pi \lambda \nonumber
\\ && \psi(0) = \xi, \quad \psi(1) = 2 \pi \lambda \cos \theta_0
\left(\frac{A}{C} - 1 \right) - \frac{F L^2}{2 \pi \lambda C} + \xi.
\end{eqnarray}
The helical solutions, $\Theta_\lambda$ in \eqref{eq:55}, are equilibria
of the rod-energy \eqref{eq:53}, subject to the boundary conditions
\eqref{eq:57}. We compute the second variation of the rod-energy 
\eqref{eq:53} about $\Theta_\lambda$ as shown below. We consider 
perturbations of the form
\begin{eqnarray}
 \label{eq:58}
 && \theta_\eps(s) =  \theta_0 + \eps \alpha(s) \nonumber \\
 && \phi_\eps (s) = 2\pi \lambda s + \eps\beta(s) \nonumber \\
 && \psi_\eps (s) = \left[ 2\pi\lambda\cos \theta_0\left( \frac{A}{C}-1 \right)
 - \frac{FL^2}{2\pi \lambda C} \right] s + \xi + \eps \gamma(s)
\end{eqnarray} with
\begin{eqnarray}
 \label{eq:59} &&
 \alpha (0) = \alpha(1) = 0 \nonumber \\
 && \beta(0) = \beta(1) = 0 \nonumber \\ && \gamma(0) = \gamma(1) = 0
\end{eqnarray} in accordance with the imposed Dirichlet conditions 
for the Euler angles. One can check that
\begin{eqnarray} \label{eq:60}
 && \frac{d^2}{d \eps^2} V[ \theta_\eps, \phi_\eps, \psi_\eps]|_{\eps = 0} =
 \int_{0}^{1} A\left\{ \alpha_s^2 + \beta_s^2 \sin^2\theta_0 
 + 4\pi\lambda \alpha \beta_s \sin\theta_0\cos\theta_0 
 - 4\pi^2\lambda^2\alpha^2\sin^2\theta_0 \right\} 
 + \frac{F L^2}{\pi \lambda} \alpha \beta_s \sin \theta_0 ~ ds 
 + \nonumber \\ && + C\int_{0}^{1} \left(\gamma_s + \beta_s\cos\theta_0 
 - 2\pi \lambda \alpha \sin\theta_0 \right)^2~ds.
\end{eqnarray}

\begin{proposition}
\label{prop:8}
The helical solutions, defined in \eqref{eq:55}, are locally stable for 
\begin{equation}
\label{eq:h3}
 A > \frac{|F|L^2}{2 \pi \lambda}
 \end{equation} and for applied forces
\begin{equation}
\label{eq:61}
\frac{|F|L^2}{2 \pi \lambda} \left( \frac{1}{A} 
+ \frac{ 4 \pi^2 \lambda^2}{ A - \frac{ |F|L^2}{2 \pi \lambda}} \right)
< \pi^2 \left(1 - 4 \lambda^2 \right).
\end{equation}
If $F=0$, then $\Theta_\lambda$ is stable for
\begin{equation}
\label{eq:62}
-\frac{1}{2} < \lambda < \frac{1}{2}.
\end{equation}

The helical solutions, defined in \eqref{eq:55}, are unstable for applied forces
\begin{equation}
\label{eq:63}
F L^2 \cos \theta_0 > 2 A \pi^2 \left(1 - \lambda^2 \right).
\end{equation}
If $F=0$, then $\Theta_\lambda$ is unstable for
\begin{equation}
\label{eq:63}
\lambda < -1 \quad \lambda > 1.
\end{equation}
\end{proposition}

\textit{Proof:} We start with the expression for the second 
variation in \eqref{eq:60}. We first note that
\begin{equation}
\label{eq:h1}
\int_{0}^{1}\frac{F L^2}{\pi \lambda} \alpha \beta_s \sin \theta_0~ds
\geq - \frac{|F|L^2}{2 \pi \lambda} \left\{ \int_{0}^{1} \alpha^2~ ds
+ \int_{0}^{1} \beta_s^2 \sin^2 \theta_0~ ds \right\}.
\end{equation}

The second variation is bounded from below by
\begin{eqnarray} \label{eq:64}
  && \frac{d^2}{d \eps^2}V[ \theta_\eps, \phi_\eps, \psi_\eps]|_{\eps = 0} 
  \geq  \int_{0}^{1} A \left\{ \alpha_s^2 - 4\pi^2 \lambda^2 \alpha^2 \right\}
  - \frac{|F|L^2}{2 \pi \lambda} \alpha^2 ~ds + \nonumber \\ && 
  +  \int_{0}^{1} \left( A - \frac{|F|L^2}{2 \pi \lambda} \right) \beta_s^2 \sin^2 \theta_0 
  + 4 \pi \lambda A \alpha \beta_s \sin\theta_0 \cos\theta_0 
  + 4 \pi^2 \lambda^2 A \alpha^2 \cos^2 \theta_0~ds.
\end{eqnarray} 
It suffices to note that for $ A > \frac{|F|L^2}{2 \pi \lambda}$,
\begin{eqnarray}
\label{eq:h2}
&& \left( A - \frac{|F|L^2}{2 \pi \lambda} \right) \beta_s^2 \sin^2 \theta_0 
+ 4 \pi \lambda A \alpha \beta_s \sin\theta_0 \cos\theta_0 
+ 4 \pi^2 \lambda^2 A \alpha^2 \cos^2 \theta_0 = \nonumber \\ && = 
\left( \left( A - \frac{|F|L^2}{2 \pi \lambda} \right)^{1/2} \beta_s \sin \theta_0 
+ \frac{ 2 \pi \lambda A}{\left( A - \frac{|F|L^2}{2 \pi \lambda} \right)^{1/2}}
\alpha \cos \theta_0 \right)^2  - \frac{ 2 \pi A \lambda}{A
- \frac{|F|L^2}{2 \pi \lambda}}|F|L^2 \alpha^2 \cos^2 \theta_0.
\end{eqnarray}
Then
\begin{eqnarray}
\label{eq:h4}
&& \frac{d^2}{d \eps^2}V[ \theta_\eps, \phi_\eps, \psi_\eps]|_{\eps = 0} \geq 
\nonumber \\ && \int_{0}^{1} A \left[ \alpha_s^2 - 4 \pi^2 \lambda^2 \alpha^2
- \frac{|F|L^2}{2 \pi \lambda A}\alpha^2
-  \frac{ 2 \pi  \lambda}{A - \frac{|F|L^2}{2 \pi \lambda}}|F|L^2 \alpha^2 \right]~ ds.
\end{eqnarray}
Finally, we recall Wirtinger's inequality 
\begin{eqnarray}
\label{eq:65}
\int_{0}^{1} \alpha_s^2~ds \geq \pi^2 \int_{0}^{1} \alpha^2(s)~ds,
\end{eqnarray}
since $\alpha$ vanishes at the end-points, $s=0$ and $s=1$. 
Substituting \eqref{eq:65} into \eqref{eq:h4}, we obtain
\begin{eqnarray}
\label{eq:67}
 && \frac{d^2}{d \eps^2} V[ \theta_\eps, \phi_\eps, \psi_\eps]|_{\eps = 0} 
 \geq  \int_{0}^{1} A \left[\pi^2  - 4 \pi^2 \lambda^2  - \frac{|F|L^2}{2 \pi \lambda A}
 -  \frac{ 2 \pi  \lambda}{A - \frac{|F|L^2}{2 \pi \lambda}}|F|L^2  \right] \alpha^2 ~ ds
 \end{eqnarray}
 and it is clear that
 $$ \frac{d^2}{d \eps^2} V[ \theta_\eps, \phi_\eps, \psi_\eps]|_{\eps = 0} > 0$$
 if $$ \frac{|F|L^2}{2 \pi \lambda} \left( \frac{1}{A}
 + \frac{ 4 \pi^2 \lambda^2}{ A - \frac{ |F|L^2}{2 \pi \lambda}} \right)
 < \pi^2 \left(1 - 4 \lambda^2 \right). $$ This condition is clearly satified 
 for $|F|L^2$ sufficiently small i.e. there exists a range of tensile and
 compressive forces for which the helical equilibria in \eqref{eq:55} are
 stable in the static sense.

 \noindent{\textbf{Instability result:} Let
\begin{eqnarray}
\label{eq:68}
&& \alpha(s) = \sin 2 \pi s
\nonumber \\ && \beta(s) = \frac{\lambda \cos \theta_0}{\sin \theta_0} 
\left( \cos 2 \pi s - 1 \right) \nonumber \\ &&
\gamma(s) = \frac{\lambda}{\sin \theta_0} \left( 1 - \cos 2 \pi s \right).
\end{eqnarray} Straightforward computations shows that
\begin{eqnarray}
\label{eq:69}
&& \beta_s \sin \theta_0 + 2\pi \lambda \alpha \cos \theta_0 = 0 \nonumber \\ 
&& \gamma_s = 2 \pi \lambda \frac{\alpha}{\sin \theta_0} 
= 2\pi \lambda \alpha \sin \theta_0 - \beta_s \cos \theta_0.
\end{eqnarray}
Therefore, the second variation \eqref{eq:60}, evaluated for this 
choice of $(\alpha, \beta, \gamma)$, is given by
\begin{eqnarray}
\label{eq:70}
&& \frac{d^2}{d \eps^2} V[ \theta_\eps, \phi_\eps, \psi_\eps]|_{\eps = 0} 
= \int_{0}^{1} A \left\{ \alpha_s^2 - 4\pi^2 \lambda^2 \alpha^2\right\} 
- FL^2\cos\theta_0  ~ds =  \nonumber \\
&& =  2 A \pi^2  \left(1 - \lambda^2 \right) - FL^2 \cos\theta_0 
\end{eqnarray} and the second variation is negative if
$$ F L^2 \cos \theta_0 > 2 A \pi^2 \left(1 - \lambda^2 \right). $$
This completes the proof of Proposition~\ref{prop:8}. $\Box$

\begin{figure}
\begin{center}
\includegraphics[scale = 0.4]{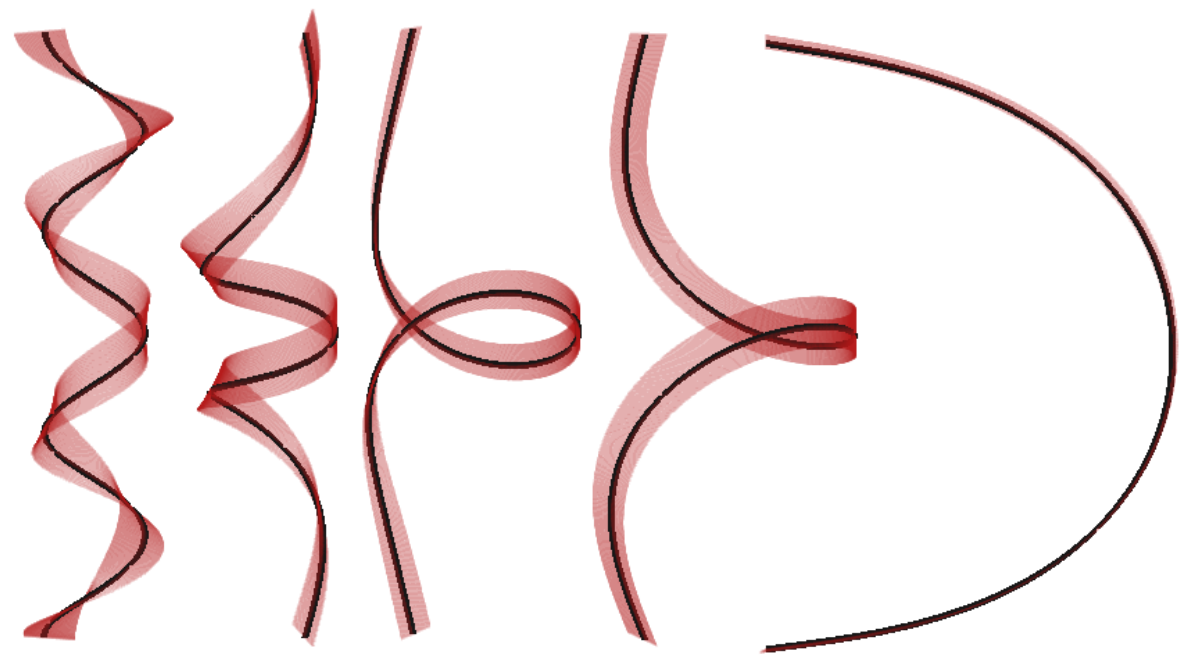}
\end{center}
\caption{Evolution of an unstable helix under a $L^2$-gradient flow. Parameters
are $L=1$, $C/A=3/4$, $\lambda = 1$ and $F=0$. We have neumann boundary conditions 
for the Euler angles and isoperimetric constraints ensure that the endpoints
of the rod stay fixed during the evolution. We note that our theory does not
cover this experiment since we do not work with isoperimetric constraints.}
\label{P:helix_neumann}
\end{figure}

\begin{figure}
\begin{center}
\includegraphics[scale = 0.4]{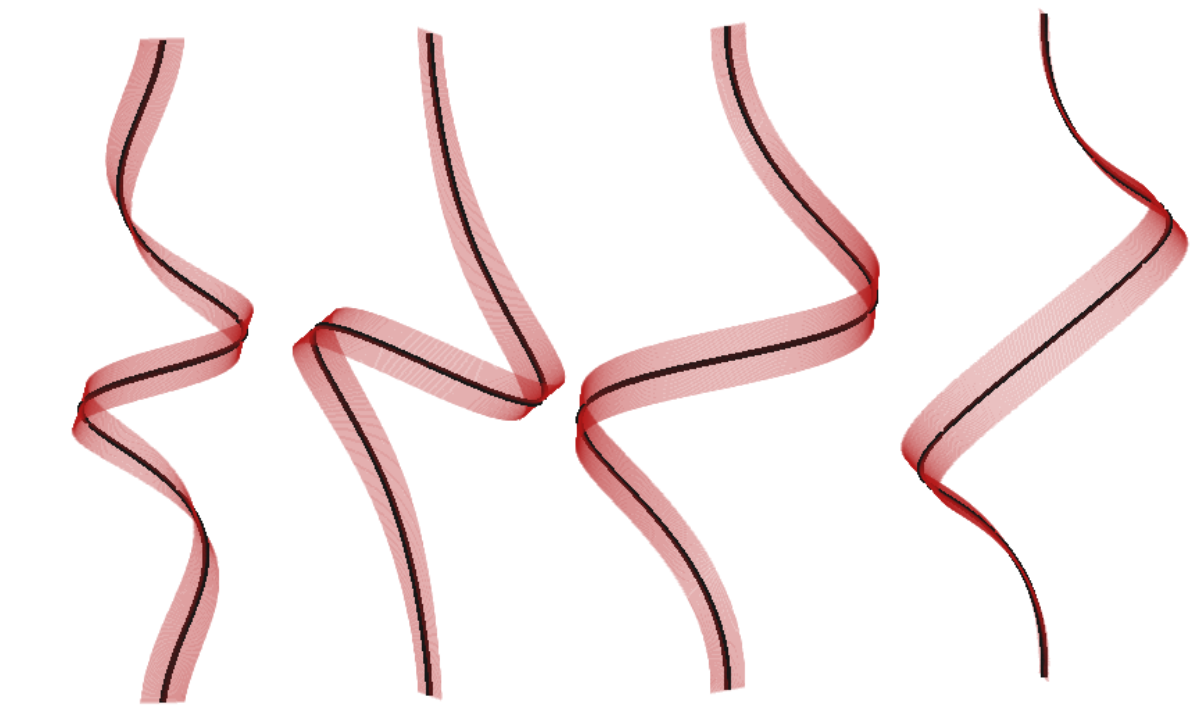}
\end{center}
\caption{Evolution of an unstable helix under a $L^2$-gradient flow. Parameters
are $L=1$, $C/A=3/4$, $\lambda=1$ and $F=0$. We have neumann boundary conditions for
$\theta$ and $\phi$ and dirichlet boundary conditions for $\psi$. Furthermore,
isoperimetric constraints ensure that the endpoints of the rod stay fixed during 
the evolution. As in Figure~\ref{P:helix_neumann} we note, that our theory does 
not cover this experiment.}
\label{P:helix_mixed}
\end{figure}

\section{Localized buckling solutions}
\label{sec:goriely}

Next, we look at nontrivial solutions of the Euler-Lagrange 
equations derived in~\cite{NiGo1999}. It is shown that 
$\Theta: \mathbb{R} \to \mathbb{R}^3$ defined by

\[
 \theta(s) = {\rm arccos}\Big(1 - \frac{2}{1+\tau^2}{\rm sech}^2(s)\Big)
\]
\[
 \phi(s) = {\rm arctan}\Big(\frac{1}{\tau}{\rm tanh} s\Big) + \tau s,
\]
\[
 \psi(s) = {\rm arctan}\Big(\frac{1}{\tau}{\rm tanh} s\Big) + \Big(3-\frac{2}{b}\Big)\tau s,
\]

are solutions to the Euler-Lagrange equations where the force 
$\mathbf{F}$ is given by $\mathbf{F} = \Big(0,0,\frac{2}{1-z_1}\Big)$
and $z_1 = \frac{\tau^2-1}{\tau^2+1}$. We fix a length $L$ and consider
$\Theta|_{(-L,L)}:\mathbb{R}\to\mathbb{R}^3$ being a solution of the 
Euler Lagrange equation subject to it's fulfilling boundary conditions.\\

For a given set of angles 
$U = (\zeta,\eta,\xi):\mathbb{R}\to\mathbb{R}^3$ we define 
the rod $\mathbf{r}_U:\mathbb{R}\to\mathbb{R}^3$ to be

\[
 \mathbf{r}_U(s):= 
 \int_0^s (\sin\zeta\cos\eta,\sin\zeta\sin\eta,\cos\zeta)~\mbox{d}\tilde{s},
\]

i.e., we solve $\frac{d \mathbf{r}_U}{d s}=\mathbf{d}_{3,U}$.
For a set of parameters $\tau = 1/2, 1, 2$, we will discuss numerically 
the stability of these localized buckling solutions. In order to compare 
$\Theta$ with perturbations $\Theta_{\varepsilon}$, we
define the space 

\[
 X_{\Theta} := \{ U:(-L,L)\to\mathbb{R}^3\,:\, r_U(-L)= r_{lb}(-L), r_U(L)=r_{lb}(L),\,\, 
                                              U(-L) = \Theta(-L)\mbox{  and  }U(L) = \Theta(L)\}.
\]

The first variation of the rod-energy $V$ on $X_{\Theta}$, evaluated at $\Theta$ vanishes, so that we can
compute a perturbation of $\Theta$ with lower energy if the second variation of $V$ on $X_{lb}$ admits
a negative eigenvalue. We take the corresponding eigenfunction $U_{\Theta}$ 
and have that $V(\Theta + \varepsilon U_{\Theta}) < V(\Theta)$ for $\varepsilon>0$ sufficiently
small. This has been done numerically and the result can be seen in Figure~\ref{P:instability}.\\

Note, that for a perturbation $\Theta_{\varepsilon}:=\Theta+\varepsilon U_\Theta$ it holds
$|r_{\Theta_\varepsilon}(L) - r_\Theta(L)| \sim \varepsilon^2$, so that 
$\Theta_\varepsilon\not\in X_\Theta$. That means, $\Theta_\varepsilon$ has lower energy but
isoperimetric constraints are only satisfied up to order $\varepsilon^2$. In a second set of
experiments we use a $L^2$ gradient flow as derived in Section~\ref{sec:numerics} and fix the
endpoints so that $r_{\Theta(t)}(L) = r_{\Theta(0)}(L)$ for $t>0$. For $\tau=1/2,1,2$ we
plot the decay of energy in Figure~\ref{P:ener_loc_buckling} and deduce that the localizsed
buckling solutions are unstable.

\begin{figure}
\begin{center}
\includegraphics[scale = 0.4]{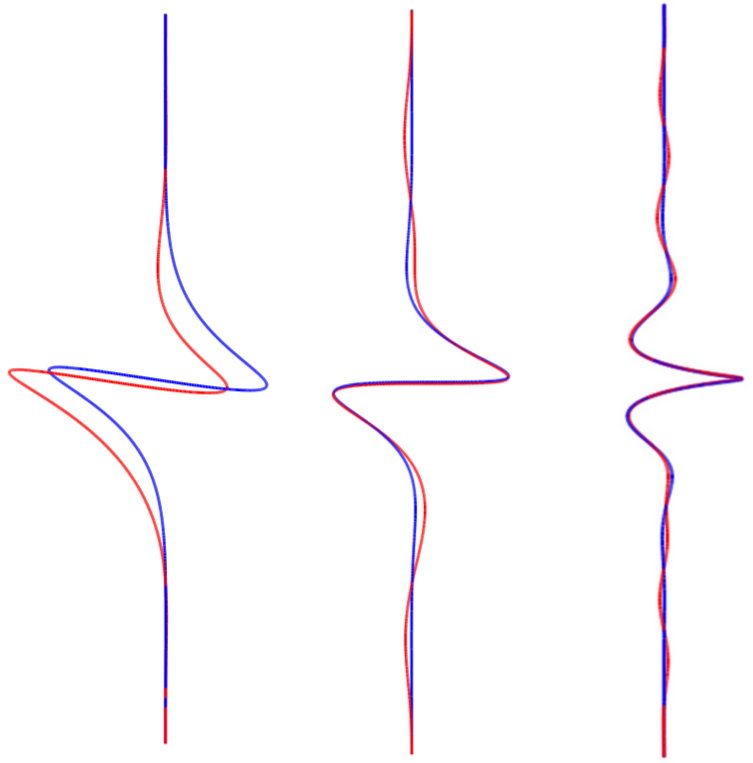}
\end{center}
\caption{Local buckling solutions (blue) and perturbations with compact support (red). We set $L=10$
and $C/A=3/4$. From left to right the solutions correspond to $\tau = 1/2$, $\tau=1$ and $\tau=2$.
The minimal eigenvalues of the second derivative of the rod-energy was $\lambda_{min} = -106.69$,.
$\lambda_{min} =  -189.50$ and $\lambda_{min} = -489.80$ for the three parameters, respectively.}
\label{P:instability}
\end{figure}

\begin{figure}
\begin{center}
\includegraphics[scale = 0.15]{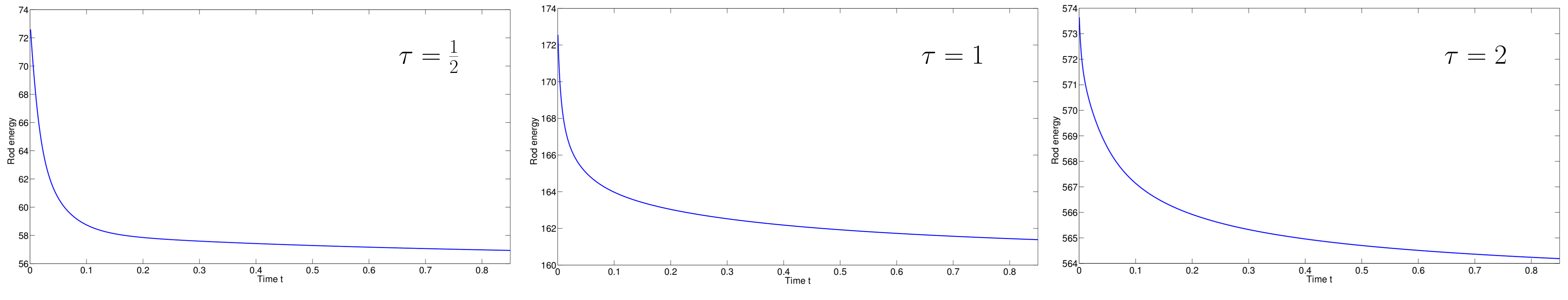}
\end{center}
\caption{Evolution under a $L^2$ gradient flow: Initial data for the 
gradient flow are the local buckling solutions with parameter $\tau = \frac{1}{2},1,2$. We plot the decay of energy
during the evolution with fixed endpoints and Dirichlet boundary conditions for all three angles.}
\label{P:ener_loc_buckling}
\end{figure}

\begin{figure}
\begin{center}
\includegraphics[scale = 0.4]{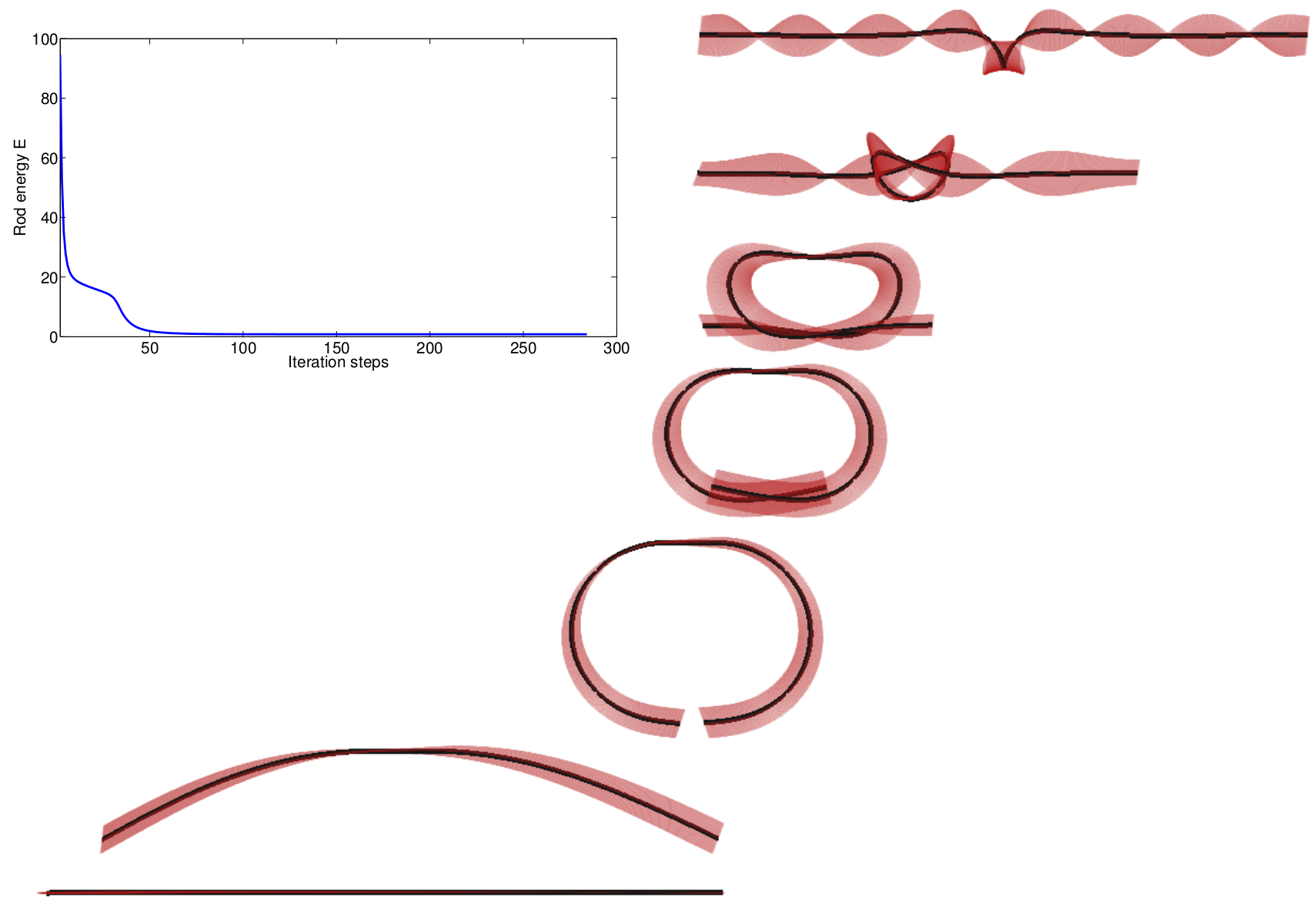}
\end{center}
\caption{Evolution of a local buckling solution under the $L^2$-gradient flow. 
Parameters are $L = 10$, $C/A=3/4$ and $\tau=1$. Isoperimetric constraints 
ensure that $x$ and $y$ components of the endpoints are fixed during the evolution. 
On the upper left we plot the decay of energy during the evolution.}
\label{P:localbuckling_neumann}
\end{figure}

\section{Numerics}
\label{sec:numerics}

Let $I=[-L,L]$, $h=(2L)/N$ be a grid size and $(-L=s_0,s_1,\dots,s_N=L)$ a uniform partition 
of $I$ with nodes $s_i = -L + ih$, $i=0,\dots,N$, $N\in\mathbb{N}$.
We define $\mathcal{S}^1(I)$ to be the space of piecewise affine, globally 
continuous functions and $\mathcal{S}_0^1(I) = \{u^h\in\mathcal{S}^1(I)\,:\, u^h(-L) = u^h(L) = 0\}$. 
Then, every $u\in\mathcal{S}^1(I)$ is clearly defined by its nodal values $(u(s_0),\dots,u(s_N))$ 
so that we can identify $\mathcal{S}^1(I)$ with $\mathbb{R}^{N+1}$. The discrete energy 
$V^h:\mathbb{R}^{3(N+1)} \to \mathbb{R}$ is defined as

\[
 (\theta^h,\phi^h,\psi^h) \mapsto \int_{0}^{1} \frac{A}{2}\left( (\theta^h_s)^2 +
(\phi^h_s)^2 \sin^2\theta^h \right) + \frac{C}{2} \left(\psi^h_s +
\phi^h_s\cos\theta^h \right)^2 + \mathbf{F}L^2\cdot \dvec_3^h~ ds,
\]

where $\dvec_3^h = (\sin\theta^h\cos\phi^h,\sin\theta^h\sin\phi^h,\cos\theta^h)$. We can compute the first variation of $V^h$ 
in a straightforward manner, 

\[
 \langle \delta V^h(\Theta^h), (\alpha^h,\beta^h,\gamma^h)\rangle = \frac{\mbox{d}}{\mbox{d}\epsilon}|_{\epsilon=0} 
 V^h(\theta^h+\epsilon \alpha^h,\phi^h+\epsilon \beta^h,\psi^h+\epsilon \gamma^h).
\]

The semi-discrete in space $L^2$-gradient flow is then defined as

\[
 \Big(\partial_t\Theta^h, (\alpha^h,\beta^h,\gamma^h)\Big)_{L^2} =  - \langle \delta V^h(\Theta^h), (\alpha^h,\beta^h,\gamma^h)\rangle,
\]

where $(u,v)_{L^2} = \int_{-L}^L u\cdot v\mbox{d}s$ is the standard $L^2$-inner product. Given a time step size
$\kappa>0$ we define the time steps $t_j = \kappa j$. An implicite time-discretization
of the $L^2$-gradient flow leads to a family of angles $(\Theta_j^h)_{j\in\mathbb{N}}$ related to the time step
$t_j$. Given $\Theta_j^h \in \mathbb{R}^{3(N+1)}$, the angles at time $t_j$, we compute the discrete velocity
$d_t\Theta^h_{j}$ as a solution of

\[
 \Big(d_t\Theta^h_{j},(\alpha^h,\beta^h,\gamma^h)\Big) 
 = - \langle \delta V^h(\Theta^{h}_{j} + \kappa d_t\Theta^h_{j}), (\alpha^h,\beta^h,\gamma^h)\rangle,
\]

and update $\Theta^{h}_{j+1} = \Theta^{h}_{j} +  \kappa d_t\Theta^h_{j}$.

\subsection{Isoperimetric Constraints}

We now introduce a method for the conservation of isoperimetric constraints during the evolution. 
The idea goes back to~\cite{BoNoPa2010,BDNR2012} where it was used to ensure conservation of 
area and mass of biomembranes during a similar energy minimization procedure. We will focus on the constraint

\[
 x(0) = x(1),\quad \mbox{i.e.} \quad \int_{-L}^L {\rm sin}\,\theta\,{\rm cos}\,\phi\,\mbox{d}s = 0,
\]

and note, that adding more side conditions is straightforward. We introduce the extended energy 

\[
 W^h(\Theta^h) = V^h(\Theta^h) + \lambda\left(\int_{-L}^L{\rm sin}\,\theta^h\,{\rm cos}\,\phi^h\,\mbox{d}s\right).
\]

and compute the first variation with respect to $\Theta^h$

\[
 \langle\delta W^h(\Theta^h),(\alpha^h,\beta^h,\gamma^h)\rangle = 
 \langle\delta V^h(\Theta^h),(\alpha^h,\beta^h,\gamma^h)\rangle + 
 \lambda\left(\int_{-L}^L\left(\alpha^h{\rm cos}\,\theta^h\,{\rm cos}\,\phi^h-\beta^h{\rm cos}\,\theta^h\,{\rm sin}\,\phi^h \right)\,\mbox{d}s\right).
\]

Following~\cite{BoNoPa2010} we compute in each time-step the velocities $v_{V^h}$ and $v_{Iso}$ via

\[
 \Big(v_{V^h}, (\alpha^h,\beta^h,\gamma^h)\Big) =  - \langle \delta V^h(\Theta^h), (\alpha^h,\beta^h,\gamma^h)\rangle,
\]

and

\[
 \Big(v_{Iso}, (\alpha^h,\beta^h,\gamma^h)\Big) =  - \int_{-L}^L\left(\alpha^h{\rm cos}\,\theta^h\,{\rm cos}\,
 \phi^h-\beta^h{\rm cos}\,\theta^h\,{\rm sin}\,\phi^h \right)\,\mbox{d}s
\]

and define the function

\[
 \rho_j(\lambda) = \int_{-L}^L{\rm sin}\,\theta^h(\lambda)\,{\rm cos}\,\phi^h(\lambda)\,\mbox{d}s, 
\]

where $\Theta(\lambda) = \Theta^h_j + \kappa(v_{V^h} + \lambda v_{Iso})$. Now, we use a Newton iteration
to compute a solution $\lambda_j$ of $\rho^j(\lambda) = 0$ and set 
$\Theta^h_{j+1} = \Theta^h_j + \kappa\left( v_{V^h} + \lambda_j v_{Iso} \right)$.\\

{\bf Fully discrete gradient flow with constraints.} Given a tolerance $TOL>0$, 
a grid size $h>0$ and a partition of $[-L,L]$, we start with an initial set of
angles $\Theta_0^h$ and time-step size $\kappa>0$. We set $j:=0$ and iterate on 
$j$ the following steps:

\begin{itemize}
 \item[(1)] Compute $v_{V^h}, v_{Iso}\in[\mathcal{S}_0^1(\mathcal{T})]^3$ satisfying 
             \[
              \Big(v_{V^h}, (\alpha^h,\beta^h,\gamma^h)\Big) =  - \langle \delta V^h(\Theta^h), (\alpha^h,\beta^h,\gamma^h)\rangle,
             \]
            and
             \[
              \Big(v_{Iso}, (\alpha^h,\beta^h,\gamma^h)\Big) =  - \int_{-L}^L\left(\alpha^h{\rm cos}\,\theta^h\,{\rm cos}\,\phi^h-\beta^h{\rm cos}\,
              \theta^h\,{\rm sin}\,\phi^h \right)\,\mbox{d}s,
             \]
            for all $(\alpha^h,\beta^h,\gamma^h)\in[\mathcal{S}^1_0(\mathcal{T})]^3$.
 \item[(2)] Compute a solution $\lambda_j$ of $\rho^j(\lambda) = 0$ and set
            $\Theta^h_{j+1} = \Theta^h_j + \kappa\left( v_{V^h} + \lambda_j v_{Iso} \right)$.
 \item[(3)] Stop if ${\rm res}^j = \left|V^h(\Theta^h_{j+1}) - V^h(\Theta^h_j)\right| < TOL$. Otherwise set $j=j+1$ and go to~(1).
\end{itemize}

\section{Conclusions}
\label{sec:conclusions}

In this paper, we study three different types of rod equilibria, including both trivial and buckled solutions,
in a fully 3D setting with different types of boundary conditions and isoperimetric constraints. The analytic methods
in Section~\ref{sec:straight} and Section~\ref{sec:helices} are relatively explicit and transparent, only depending on
integral inequalities. These methods yield explicit stability estimates in terms of the twist, load and elastic
constants and give valuable information about the incipient instabilities, as illustrated in Section~\ref{sec:bifurcation}. 
In particular, we bypass the traditional problems with Neumann boundary conditions in Section~\ref{sec:neumann}.
The numerical experiments in Sections~\ref{sec:helices} and \ref{sec:goriely} could be carried out systematically
to devise model conditions under which these non-trivial solutions could be stabilized.
The work in this paper is only foundational for 3D studies of rod equilibria and there are several open
directions e.g. dynamical studies of the fully nonlinear Kirchhoff rod equations, inclusion of intrinsic 
curvature into the stability analysis, non-equilibria transitions between different equilibria as a function of 
the external load and the role of external loads in stabilization and destabilization effects. However, the analytic
methods in this paper can be carried over to more complicated situations of extensible-shearable rods or rods with 
intrinsic curvature and the numerical methods can be readily adapted to include topological and various boundary constraints.
Therefore, we believe that these methods provide new tools and approaches to applied mathematicians in this area and we hope
to report on new 3D effects in future work.

\noindent \textbf{Acknowledgments:}

AM and AR thank Alain Goriely for several helpful discussions and 
suggestions, which led to the improvement of this manuscript. 
AM and AR also thank John Maddocks, Sebastien Neukrich and 
Gert van der Heijden for helpful comments. AM is supported by an 
EPSRC Career Acceleration Fellowship EP/J001686/1, an OCCAM Visiting 
Fellowship and a Keble Research Grant. AR is supported by
KAUST, Award No. KUK-C1-013-04 and the John Fell OUP fund.

\bibliographystyle{acm}
\bibliography{rods}
\end{document}